\documentclass[%
reprint, nature
]{revtex4-2}

\usepackage{graphicx}% Include figure files
\usepackage{amsmath,amssymb}
\usepackage{xfrac}
\usepackage[separate-uncertainty=true,multi-part-units=single]{siunitx}
\usepackage[english]{babel}
\usepackage{relsize}
\usepackage{float}
\usepackage{hyperref}
\usepackage{braket}
\usepackage{cleveref}
\usepackage{placeins}

\begin{document}

\title[Programmable Bell State Generation in an Integrated Thin Film Lithium Niobate Circuit]{Programmable Bell State Generation in an Integrated Thin Film Lithium Niobate Circuit}


\author{Andreas Maeder}
\thanks{This author contributed equally to this work.}
\email{maederan@phys.ethz.ch}
\author{Robert J. Chapman}
\thanks{This author contributed equally to this work.}

\author{Alessandra Sabatti}

\author{Giovanni Finco}

\author{Jost Kellner}

\author{Rachel Grange}
\affiliation{ETH Zurich, Department of Physics, Institute for Quantum Electronics, Optical Nanomaterial Group, Zurich, Switzerland}

\begin{abstract}
	Entanglement is central to quantum technologies such as cryptography, sensing, and computing. Photon pairs generated via nonlinear optical processes are excellent for preparing entangled states due to their long coherence times and compatibility with fiber optic networks. Steady progress in nanofabrication has positioned lithium niobate-on-insulator (LNOI) as a leading platform for monolithic integration of photon pair sources into optical circuits, leveraging its strong second-order nonlinearity. Here, we present a reconfigurable photonic integrated circuit on LNOI, which combines two on-chip photon pair sources with programmable interferometers, enabling generation of entangled states. The pair sources achieve a source brightness of $\SI{26}{\mega\hertz\per\nano\meter\per\milli\watt}$ while maintaining a coincidence-to-accidental ratio above 100. We successfully interfere the two sources with \SI{99.0 \pm 0.7}{\percent} visibility, demonstrating the indistinguishability required for producing entanglement on-chip. We show preparation of any of the maximally entangled Bell states with fidelity above \SI{90}{\percent} verified by quantum state tomography. These results establish LNOI as a compelling, scalable platform to explore integrated quantum photonic technologies enabled by high-brightness sources of entangled quantum states. 
\end{abstract}

\maketitle

\section*{Introduction}

Photonic systems are at the forefront of research in quantum technologies for applications like quantum computing, quantum cryptography or quantum sensing~\cite{obrienPhotonicQuantumTechnologies2009}. These applications benefit from the intrinsic advantages of photons, such as their low decoherence, ease of long distance transmission through optical fiber and weak coupling to the environment. Entanglement is the key requirement for quantum error correction, quantum secure communication, and quantum computational advantage for classically intractable tasks like boson sampling. Therefore, a scalable, efficient, stable and compact source of entangled photons is one of the key requirements for advancing quantum information science.

Fundamental demonstrations of quantum optics and its applications in technology have been based on free-space or fiber-optic systems, which use spontaneous parametric downconversion (SPDC) as a source of entangled photons. This process relies on the second order nonlinearity ($\chi^{(2)}$) of non-centrosymmetric crystals to generate photon pairs entangled in polarization, frequency, space or time~\cite{meraner2021ApproachingTsirelsonBound,olislagerFrequencyBinEntangledPhotons2010,krennEntanglementByPathIdentity,MarcikicDistributionTimebin2004}. Among the most common $\chi^{(2)}$-crystals is lithium niobate (LN) which was used for experimental realizations of high fidelity entangled states~\cite{KimPhaseStableSource2006}, generation of squeezed light \cite{LenziniIntegratedPhotonicPlatform2018}, quantum key distribution \cite{tanzilliPPLNWaveguideQuantum2002} or quantum teleportation \cite{bussieresQuantumTeleportationTelecomwavelength2014}. 
Although these LN pair sources offer the required high brightness, the experiments require phase stabilization and bulky optical elements for transformation of the generated quantum state, which makes the approach inherently non-scalable.

Integrated photonics is a scalable solution for quantum photonic technologies, with demonstrations of unprecedented complexity in recent years~\cite{maringVersatileSinglephotonbasedQuantum2024,aghaeeradScalingNetworkingModular2025}. Unlike free-space implementations, which commonly exploit entanglement in the polarization degree of freedom, integrated photonics predominantly relies on spatial encoding of photons. In this encoding, quantum operations can be realized using on-chip programmable interferometers \cite{knillSchemeEfficientQuantum2001}, as has been demonstrated for example in silicon, silicon nitride or silica photonics \cite{wangMultidimensionalQuantumEntanglement2018,Carolan2015,taballione20ModeUniversalQuantum2023,ThompsonIntegratedWaveguideCircuits2011}. 

The development of lithium niobate-on-insulator (LNOI) wafers enables the combination of the strong $\chi^{(2)}$ nonlinearity of LN with low-loss optical waveguides that support tight bending radii and scalable fabrication processes \cite{lukeWaferscaleLowlossLithium2020}, making it an ideal platform for exploring integrated quantum photonics. Numerous static and dynamic building blocks for classical applications have been realized on this platform~\cite{zhuIntegratedPhotonicsThinfilm2021,pohl100gbdWaveguideBragg2021,lomonteScalableEfficientGrating2024,prencipeTunableUltranarrowbandGrating2021}. For quantum photonics, realizations of high brightness sources~\cite{zhaoHighQualityEntangled2020}, on-chip two-photon interference~\cite{babelDemonstrationHongOuMandelInterference2023,chapmanOnchipQuantumInterference2025,maederOnchipTunableQuantum2024}, and compatibility with superconducting single photon detectors in cryogenic environments~\cite{lomonteSinglephotonDetectionCryogenic2021,sayemLithiumniobateoninsulatorWaveguideintegratedSuperconducting2020} represent essential developments. Despite these achievements, most research on LNOI has remained limited to isolated devices or simple assemblies of few elements, rather than fully integrated quantum circuits. Moreover, architectures that bring together high brightness on-chip sources and reconfigurable circuits remains unexplored in LNOI. 

\begin{figure*}[t!]
	\includegraphics[width=\textwidth]{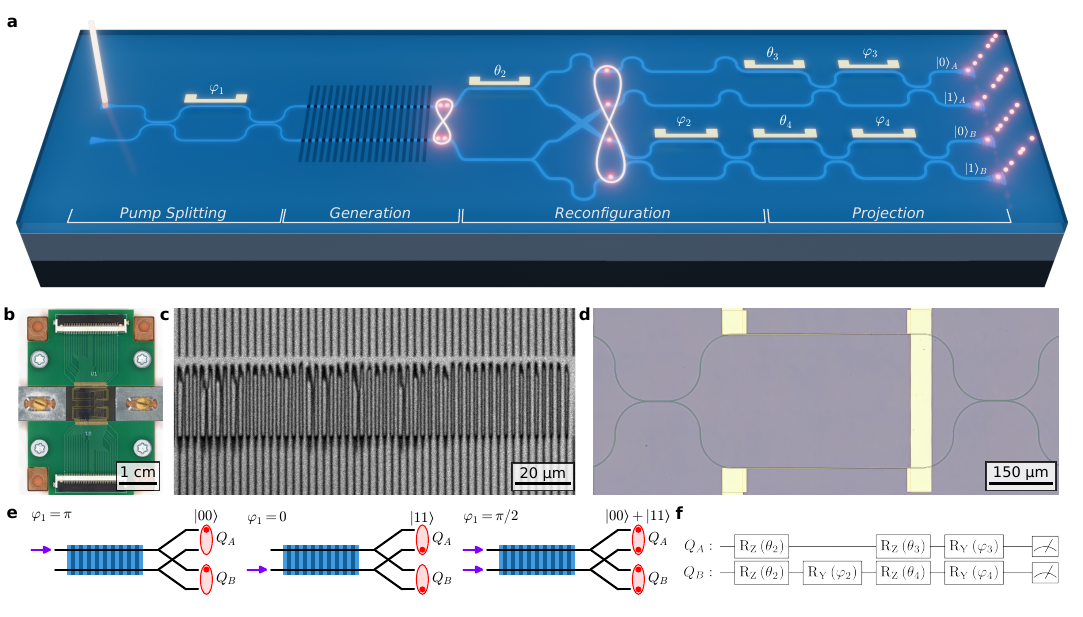}
	\caption{\textbf{Working principle and fabrication result of LNOI circuit.} (\textbf{a}) CW pump light ($\lambda_p = \SI{775}{\nano\meter}$) is coupled to the chip via grating couplers and split using an MZI. This generates a superposition of photon pairs in two periodically poled waveguides which is split into two dual-rail encoded qubits, creating an entangled Bell state. Integrated thermo-optic phase shifters and MZIs are used to transform and project the generated two-qubit state. (\textbf{b}) Image of the final photonic integrated circuit including electrical packaging. (\textbf{c}) Two-photon microscopy image of periodically poled region prior to waveguide etching. (\textbf{d}) Microscope image of an integrated MZI with TO phase shifter. (\textbf{e}) Different pumping scheme generating different states dependent on the pump phase $\varphi_1$. (\textbf{f}) Equivalent quantum circuit representation in the two-qubit picture.} \label{fig1}
\end{figure*}

This work combines high performance building blocks developed on the LNOI platform into a single, programmable circuit which generates entangled Bell states on-chip. We monolithically integrate a pair of periodically poled LNOI waveguides with a reconfigurable interferometric circuit to realize path encoded Bell states. The photon pair sources have an on-chip spectral brightness of \SI{26}{\mega\hertz\per\milli\watt\per\nano\meter} and we observe $\SI{99.0 \pm 0.7}{\percent}$ two-photon interference visibility. We verify the presence of on-chip generated states by quantum state tomography, which allow us to reconstruct the full density matrix of the state. The state projections required for quantum state tomography are performed directly on-chip, making the approach inherently phase-stable. Through additional control of phases, we are able to reconfigure the circuit to generate different two-qubit states. We demonstrate the generation of computational basis states with fidelities above \SI{95}{\percent} and entangled Bell states with fidelities above \SI{90}{\percent}. These results mark an important step toward scalable quantum photonic systems leveraging $\chi^{(2)}$-sources. By combining many building blocks into a single, programmable device, this work showcases the potential of the LNOI platform for realizing advanced quantum information processing circuits.

\section*{Results}

\subsection*{Device Principle}

\begin{figure*}[t!]
	\includegraphics[width=\textwidth]{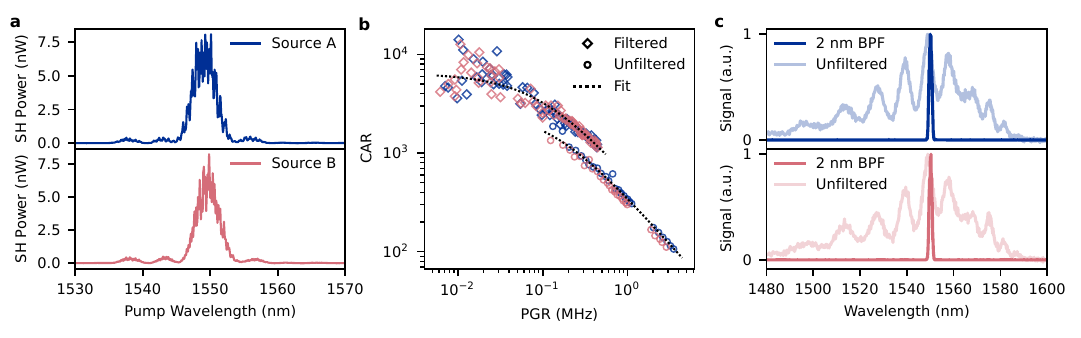}
	\caption{\textbf{Characterization measurements of calibration photon pair sources.} (\textbf{a}) Second harmonic (SH) spectra for two periodically poled waveguide, source A and B, respectively. (\textbf{b}) Coincidence-to-accidental ratio (CAR) as a function of pair generation rate (PGR) for both sources. Circle markers are measurements without spectral filtering, diamond markers are measurements including a \SI{2}{\nano\meter} bandpass filter (BPF) (\textbf{c}) Time-of-flight spectroscopy measurements of top and bottom pair source with and without the BPF.} \label{fig2}
\end{figure*}

\Cref{fig1}a shows a schematic of the LNOI circuit and an image of the fabricated and electrically packaged device is shown in \cref{fig1}b. It combines two waveguide-integrated SPDC sources enabled by local periodic poling of the LN film. This periodic inversion of the $\chi^{(2)}$ coefficient (see \cref{fig1}c) implements quasi-phase matching, compensating the momentum mismatch between pump, signal, and idler waves ensuring efficient photon pair generation~\cite{HumQuasiPhasematching2007}. By pumping both waveguides at $\lambda_p = \SI{775}{\nano\meter}$, signal and idler photons centered around the degenerate wavelength $\lambda_s = \lambda_i = \SI{1550}{\nano\meter}$ are created. Importantly, because the sources are pumped with a continuous wave (CW) laser, at any given time only a single photon pair is generated as a superposition of emission from the two sources \cite{chapmanOnchipQuantumInterference2025}. A Mach-Zehnder interferometer (MZI, see \cref{fig1}d) prior to the sources allows for controlling the relative pump power by adjusting its phase difference $\varphi_1$. After the poled waveguides, the photon pairs are probabilistically split into two dual-rail encoded qubits $Q_A$ and $Q_B$ using two Y-splitters and a waveguide crossing. As shown in \cref{fig1}e, the pumping scheme can be adjusted to produce either pure computational basis states $\ket{00}$ or $\ket{11}$ ($\varphi_1 = 0~\text{or}~\pi$) or a maximally entangled Bell state $\ket{\Phi^+} \propto \ket{00} + \ket{11}$ ($\varphi_1 = \pi/2$). Two additional phase shifters controlling $\theta_2$ and $\varphi_2$ (see \cref{fig1}a) are used to modify the generated state further. If equal pumping with $\varphi_1 = \pi/2$ and equal source efficiency $\eta_A = \eta_B = 1$ are assumed, the state can be rewritten in Bell basis as 
\begin{align}
	\begin{split}
		\ket{\psi} \propto i \sin{\left( \theta_2 \right) } \sin{ \left( \frac{\varphi_2}{2}\right)} &\ket{\Phi^+} \\
		+ \quad \cos{\left( \theta_2 \right) } \sin{ \left( \frac{\varphi_2}{2}\right)} &\ket{\Phi^-}\\
		+ \quad \cos{\left( \theta_2 \right) } \cos{ \left( \frac{\varphi_2}{2}\right)} &\ket{\Psi^+} \\
		+ \quad i \sin{\left( \theta_2 \right) } \cos{ \left( \frac{\varphi_2}{2}\right)} &\ket{\Psi^-},
	\end{split}
	 \label{eq:BellBasis}
\end{align}
where global phase factors and normalization constants have been omitted. This establishes that $\varphi_2$ dictates whether the odd or even parity Bell states, $\ket{\Psi^\pm}$ or $\ket{\Phi^\pm}$, respectively, are being generated, while $\theta_2$ controls the relative symmetry of the resulting state. Given the control over the phases through thermo-optic (TO) phase shifters, the circuit can be reconfigured to generate any one of the four Bell states. Additionally, in the case where only one source is pumped, the phases can be programmed to prepare the other two computational basis states $\ket{01}$ and $\ket{10}$. A more extensive theoretical derivation of the state generation is given in the supplementary material.

The second half of the circuit shown in \cref{fig1}a facilitates quantum state tomography to reconstruct the full density matrix of the state generated on-chip. For this, an additional MZI (phases $\varphi_3, \varphi_4$) with an external phase shifter (phases $\theta_3, \theta_4$) is added to each qubit. This applies unitary transformations to the qubits, which is used to implement the required projections, as illustrated in \cref{fig1}f with the quantum circuit diagram realized by the LNOI circuit. As for all the previous phases, the tomography phases are physically controlled using TO phase shifters (see \cref{fig1}d). Coupling of the pump laser to the circuit as well as the generated photons to a single photon detection system is facilitated through grating couplers. More details on the device fabrication can be found in the Methods section.

\subsection*{Photon Pair Sources}

The high $\chi^{(2)}$ nonlinearity of LNOI enables efficient photon pair generation via SPDC. By etching waveguides in regions where the LN film has been periodically poled, high brightness integrated photon pair and heralded single photon sources can be realized~\cite{zhaoHighQualityEntangled2020}. A two-photon microscope image of the inverted domains, acquired prior to waveguide etching, is shown in \cref{fig1}c (see Methods for fabrication details).

%While spontaneous four-wave mixing (SFWM) based sources common in silicon photonics typically use cm-long waveguides or resonant enhancement and typically require pulsed lasers~\cite{MonteleonePackagedFoundryFabricated2022,alexanderManufacturablePlatformPhotonic2025,maSiliconPhotonicEntangled2017}, a single few mm-long poled LNOI waveguide is sufficient to generate high pair rates with a continuous wave (CW) pump laser.

\begin{figure*}[t!]
	\includegraphics[width=\textwidth]{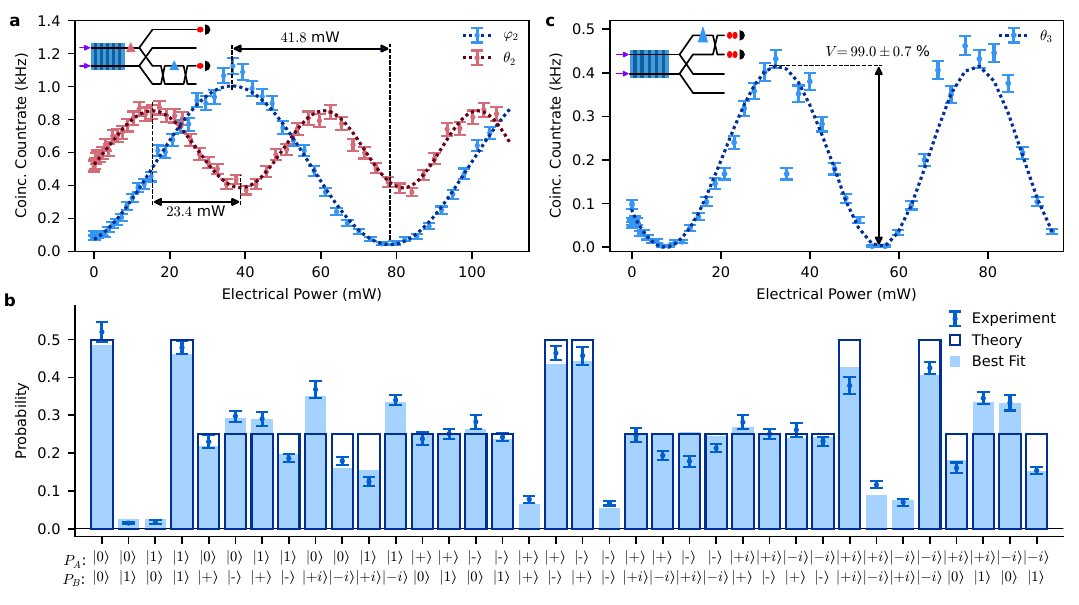}
	\caption{\textbf{Interference characterization and projection measurements.} (\textbf{a}) Calibration measurements for thermo-optic phase shifters of reconfiguration phases $\varphi_2$ and $\theta_2$. Dotted lines are calibration fits. The inset indicates simplified circuit diagram. (\textbf{b}) Overcomplete set of projections used for quantum state tomography of the $\ket{\Phi^+}$ state. Experimentally measured probability is compared to the theoretically expected results and the obtained best fit using a global minimization algorithm. (\textbf{c}) Two-photon interference measurement using pairs from independent sources with \SI{99.0 \pm 0.7}{\percent} visibility $V$. The inset represents a simplified schematic of the experiment.} \label{fig3}
\end{figure*}

For quantitative characterization of the SPDC sources, we fabricated a pair of calibration waveguides in an identical periodically poled region separate from the full circuit. \Cref{fig2}a shows the measured second harmonic signal generated by calibration source A and B. We observe very good spectral overlap between the two second harmonic intensities, and a phase matching wavelength close to \SI{1550}{\nano\meter} as targeted in the design. The normalized second harmonic conversion efficiencies of the two sources are \SI{2150}{\percent\per\watt\per\centi\meter\squared} and \SI{2708}{\percent\per\watt\per\centi\meter\squared}, respectively, which are close to the theoretical limit of \SI{3300}{\percent\per\watt\per\centi\meter\squared}.

Having established near identical phase matching and second harmonic generation efficiency, we characterize the SPDC process by probabilistically splitting the photon pairs with an off‑chip fiber-beamsplitter and performing coincidence measurements between the two outputs. We compare photon pairs which are not spectrally filtered beyond the response of the grating couplers to those passing through a \SI{2}{\nano\meter} bandpass filter. \Cref{fig2}b presents the on-chip pair generation rate (PGR) and coincidence-to-accidental ratio (CAR) for both sources. The on-chip pump powers in these measurements vary between \SI{0.1}{} and  \SI{10}{\micro\watt} for both filtered and unfiltered case. The CAR follows the expected PGR\textsuperscript{-1} dependence (dashed lines in \cref{fig2}b)~\cite{zhangHighperformanceQuantumEntanglement2021}.

The on-chip source brightness of the unfiltered sources is estimated to be \SI{1.5}{\giga\hertz\per\milli\watt} and \SI{1.7}{\giga\hertz\per\milli\watt} for source A and B, respectively. The bandpass filter reduces the source brightness to \SI{51.6}{} and \SI{51.3}{\mega\hertz \per \milli\watt}, respectively. This corresponds to a spectral brightness of approximately \SI{26}{\mega\hertz \per \milli\watt \per \nano\meter} for both sources. These values compare well to existing literature on SPDC in periodically poled LNOI waveguides~\cite{zhaoHighQualityEntangled2020}, and outperform similar photon pair sources based on third order nonlinearity in silicon waveguides~\cite{ShinPhotonPairGeneration2023,GuoHighCoincidenceAccidentalRatio2017}.

To confirm the spectral properties of the generated photons, we used a time-of-flight technique to measure the photon spectra directly (see Methods). The single photon spectra measured with and without bandpass filter are shown in \cref{fig2}c. The unfiltered photons show a broad spectrum of around \SI{100}{\nano\meter} bandwidth, which is expected for the type-$0$ SPDC process used here. It is limited by the spectral bandwidth of the grating couplers used to couple the photons to fiber. The filtered photons inherit the spectrum from the bandpass filter as expected.

Overall, these results demonstrate the high efficieny of the on-chip SPDC photon pair sources. The excellent spectral overlap and matched performance of the two sources are essential for high-visibility quantum interference, forming the foundation for the generation of entangled states.

\subsection*{Calibration and Tomography Measurements}

Accurate operation of the programmable circuit relies on calibration of the TO phase shifters. To achieve this, we determine the phase-voltage relationship of each phase shifter individually. Calibration curves for the state reconfiguration phases $\varphi_2$ and $\theta_2$ are shown in \cref{fig3}a as a function of the dissipated electrical power. A sinusoidal fit to the data provides a model of the voltage–phase relationship, which enables inversion to determine the required voltage for a target phase. Note that countrate for the $\theta_2$ calibration oscillates at roughly twice the frequency, which is consistent with the theoretical prediction in \cref{eq:BellBasis}. Each calibration is performed with other phases shifters turned off as much as possible to minimize cross-talk effects. Therefore, the single photon visibilities in \cref{fig3}a are not maximized. Moreover, for most of the calibration procedure the pump phase is set to $\varphi_1 = \pi/2$ which assumes identical source efficiencies. However, two-photon interference measurements discussed later (see \cref{fig3}c) showed that balanced pumping of the sources required $\varphi_1 \sim \pi/3$ to compensate for a mismatch in SPDC generation probability between the two sources.

\begin{figure*}[tb!]
	\includegraphics[width=\textwidth]{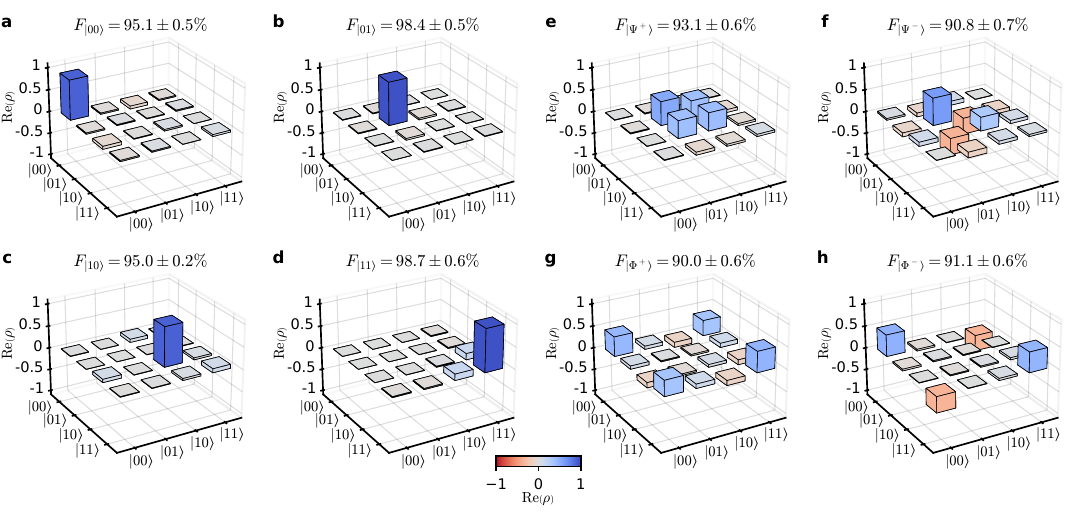}
	\caption{\textbf{Reconstructed Density Matrices.} Real part of density matrices obtained through maximum likelihood estimation for (\textbf{a}) $\ket{00}$, (\textbf{b}) $\ket{01}$, (\textbf{c}) $\ket{10}$, (\textbf{d})  $\ket{11}$, (\textbf{e}) $\ket{\Psi^+}$, (\textbf{f}) $\ket{\Psi^-}$ , (\textbf{g}) $\ket{\Phi^+}$, and (\textbf{h}) $\ket{\Phi^-}$ state. $F_{\ket{\psi}}$ indicates the fidelity of the shown density matrix to the respective target state $\ket{\psi}$. The imaginary parts are given in the supplementary material.} \label{fig4}
\end{figure*}

As mentioned in the working principle, the circuit allows for performing quantum state tomography to measure the full density matrix of generated quantum states. The tomographic reconstruction is performed using a complete set of measurements with projections onto the single-qubit Pauli eigenstates $\left\{ \ket{0}, \ket{1}, \ket{+},\ket{-}, \ket{i}, \ket{-i} \right\}$ for each qubit, yielding $6^2 = 36$ measurements. The measurements are performed by setting the projection phases to implement the qubit rotations and collecting coincidence counts over \SI{2}{\second}. This short integration time is enabled by the high brightness sources leading to an off-chip PGR of around $\SI{1}{\kilo\hertz}$ during the tomography experiments. In addition, parallel detection of all four qubit rails reduces the acquisition time to $\SI{30}{\second}$ per state tomography. We perform maximum likelihood estimation using the $36$ coincidence counts to reconstruct the full quantum state~\cite{altepeter4QubitQuantum2004}. More details on the state reconstruction can be found in the supplementary material.

To illustrate the results of the projection measurements, \cref{fig3}b shows the measured probabilities for each projection of a $\ket{\Phi^+}$ state. Already from the Pauli-$Z$ basis projections, one can identify the characteristic superposition of $\ket{00}$ and $\ket{11}$. All of the experimental results show good agreement with theoretically expected probabilities. The density matrix reconstructed via maximum likelihood estimation shows good agreement to the experimental and theoretical probabilities. These results validate the functionality of the reconfigurable on-chip projections based on TO phase shifters and support their use in more complex quantum circuit on the LNOI platform.

 %measurement stages based on TO phase shifters and MZIs on the LNOI platform.  and support their use in more complex quantum protocols as has been demonstrated in silicon photonic circuits \cite{wangMultidimensionalQuantumEntanglement2018,silverstoneQubitEntanglementRingresonator2015}.}

\subsection*{Entangled State Generation}

Stable and scalable generation of entangled photon pairs is a key requirement for photonic quantum technologies. In our integrated LNOI circuit, we use the probabilistic nature of SPDC to create path entangled states. The programmability enables the preparation of different entangled states, making it a versatile source for various quantum protocols. 

Given that the photon splitting by the Y-splitters is entirely probabilistic, we can first investigate an entangled two-mode state. It is generated when both sources are pumped, but the photon pair is not split (see inset of \cref{fig3}c). The state generated in that case, a $N00N$ state with $N=2$, can be investigated when configuring the subsequent MZI as a $50:50$ beam splitter ($\varphi_3 = \pi/2$). In this configuration we can observe a two-photon interference effect referred to as the time-reversed Hong-Ou-Mandel effect~\cite{chenDeterministicQuantumSplitter2007,chapmanOnchipQuantumInterference2025}. With the phase $\theta_3$ we can control whether the photon pairs interfere constructively or destructively, which we report in \cref{fig3}c. The observed visibility of \SI{99.0 \pm 0.7}{\percent}, obtained from the maximum of the sinusoidal fit and the minimum measured countrate, indicates that photon pairs from separate sources are in fact indistinguishable to a very high degree. 

Having shown the capability of generating entangled two-photons states with only part of the circuit, we next demonstrate on-chip generation of eight different two-qubit states. For this, we pump the circuit with $\SI{150}{\micro\watt}$-CW laser light and measure conicidence counts between the four combinations of qubit rails. During this experiment we observe an off-chip PGR of around \SI{1}{\kilo\hertz} and the CAR around $100$. Due to insertion loss of additional on-chip components, less than optimal fiber-to-chip coupling due to usage of fiber arrays instead of single fibers, these values are lower than the ones reported for the calibration source characterization in \cref{fig2}. However, they were sufficient to conduct the experiment without the need for excessive integration time.

The programmability of the circuit is first demonstrated by preparing all four computational basis states, by pumping a single source only (see \cref{fig1}e) and using $\varphi_2$ to flip the second qubit. The real parts of the reconstructed density matrices are shown in \cref{fig4}a-d. All of them show high fidelity above \SI{95}{\percent} to the respective target state. Next, both sources are pumped with $\varphi_1 = \pi/3$, to achieve equal pair generation probability in both waveguides. Reconfiguration of $\varphi_2$ and $\theta_2$ allows us to generate each of the four Bell states. Their reconstructed density matrices are shown in \cref{fig4}e-h and have fidelities of above \SI{90}{\percent}, with $\ket{\Psi^+}$ having the highest fidelity of \SI{93.1 \pm 0.6}{\percent}. Besides fidelity, we also report the concurrence $C$ of the states, which is computed from the density matrices. For the highest fidelity state $\ket{\Psi^+}$, we find $C = \SI{0.80 \pm 0.03}{}$, which is above the limit $C=1/\sqrt{2}$ required to guarantee violation of the Clauser-Horne-Shimony-Holt (CHSH) inequality~\cite{Verstraete2002}. Notably, all four Bell states satisfy this criterion. Furthermore, we compute the von Neumann entropy $S_A$ and $S_B$ of each qubit by tracing out the other qubit. Those values vary between $0.64$ and $0.69$, which match well with the theoretical value $\ln{\left(2 \right)}$ for a maximally entangled state in two dimensions. A full table of calculated state metrics is available in the supplementary material.

All of the findings show that the integrated circuit is capable of generating entangled two-photon states directly on-chip. Moreover, the inherent phase stability of integrated photonics is leveraged by including projections required for quantum state tomography on-chip as well. The ability to prepare a complete set of Bell states at high generation rates above $\SI{1}{\kilo\hertz}$ highlights the potential of this LNOI circuit as a compact, programmable source of entangled states for quantum photonic applications.

\section*{Discussion}

In conclusion, we have demonstrated a programmable monolithic LNOI photonic integrated circuit capable of on-chip generation of entangled two-photon states. By combining two independent SPDC photon pair sources, each achieving an on-chip spectral brightness of \SI{26}{\mega\hertz\per \milli \watt}, with reconfigurable interferometric elements, we first prepared $N00N$ states which show time-reversed Hong-Ou-Mandel interference with a visibility of $\SI{99.0 \pm 0.7}{\percent}$. This confirms the high degree of indistinguishability of the independent SPDC sources. Using the full circuit, we realized two-qubit states and performed tomographic reconstruction of the density matrices using on-chip projections and only a few seconds integration time per measurement. The achieved fidelities exceed \SI{95}{\percent} for computational basis states and \SI{90}{\percent} for each of the four Bell states. Analysis of concurrence and von Neumann entropy confirms that the Bell states violate CHSH inequality and are maximally entangled with highly mixed subsystems. 

Similar experiments have been realized on silicon photonics, however with lower fidelities for both computational basis states and Bell states~\cite{silverstoneQubitEntanglementRingresonator2015,santagatiSiliconPhotonicProcessor2017}. These demonstrations rely on spontaneous four wave mixing (SFWM) photon pair sources, which require usage of pulsed lasers and cm-long waveguide spirals, to operate at reasonable PGR. Other realizations of reconfigurable interferometric circuits rely on off-chip $\chi^{(2)}$ sources or cryogenically operated quantum dots for state generation, which leads to additional coupling loss limiting the achievable state generation rates \cite{SansoniPolarizationEntangledStateMeasurement2010,pontHighfidelityFourphotonGHZ2024}. While recent work has demonstrated SPDC sources in silicon nitride using all-optical poling~\cite{dalidetPerfectTwophotonInterference2022}, its source brightness remains considerably lower than achieved here, underscoring the benefit of the intrinsic $\chi^{(2)}$ nonlinearity of LN. Our experimentally achieved off-chip PGR demonstrates that integrated SPDC sources on the LNOI platform can outperform SFWM-based sources, while requiring only mm-long straight waveguides and CW pump lasers. Considering that our circuit currently suffers from photon loss at the fiber-to-chip interface, there is significant room for improvement of the PGR. This only marks a technical challenge, seeing that efficient grating couplers with below $\SI{1}{\decibel}$ loss per grating have been achieved on the LNOI platform \cite{lomonteScalableEfficientGrating2024}. Moreover, we envision to combine such circuit with waveguide integrated single photon detectors in the future, which would significantly improve the detection rates. Furthermore, the measured two-photon interference visibility is equally high with reported SFWM-based chips~\cite{wangChiptochipQuantumPhotonic2016,silverstoneQubitEntanglementRingresonator2015}, which shows that high degrees of photon indistinguishability is achievable with periodically poled waveguides on the LNOI platform. Fabricating more waveguides in the same poled region offers a direct path to scaling to multi-qubit states. We anticipate that the superior brightness will be especially advantageous, which will be critical when realizing higher photon number states like Greenberger-Horne-Zeilinger states~\cite{ChenObservationQuantumNonlocality2024,pontHighfidelityFourphotonGHZ2024}.

This work provides a first demonstration of combining high performance linear building blocks developed on the LNOI platform with on-chip SPDC sources into a single, programmable quantum photonic circuit. Continued efforts to scale LNOI fabrication are expected to further enhance the platform, supporting the development of more advanced and scalable quantum photonic circuits.

\section*{Methods}

\subsection*{Device Fabrication}
The integrated circuit was fabricated on a \SI{300}{\nano\meter} x-cut LNOI chip with \SI{4.7}{\micro\meter} silicon dioxide bottom oxide layer. First, \SI{1.5}{\milli\meter} long and \SI{30}{\micro\meter} wide regions of periodically poled LN film are created by applying high voltage pulses to comb-like electrodes. The poling period used is $\Lambda = \SI{2.87}{\micro\meter}$. Subsequently, LN waveguides are patterned using electron-beam lithography and Argon ion milling in an inductively coupled plasma reactive ion etching tool. The etch depth of \SI{200}{\nano\meter} is controlled using an end-pointing system. A cladding layer of \SI{1}{\micro\meter} silicon dioxide is deposited on the patterned chip, followed by fabrication of the gold TO electrodes using electron beam lithography and a standard double-layer liftoff process. The TO phase shifters have a low footprint of $\SI{1}{\micro\meter} \times \SI{100}{\nano\meter} \times \SI{0.4}{\milli\meter}$. For electrical connections to the phase shifters, a \SI{300}{\nano\meter} thick set of gold routing electrodes is deposited. These electrodes are wirebonded to a printed circuit board mounted together with the photonic chip on a copper mount, which is held at $\SI{22}{\celsius}$ with a Peltier element during all the measurements.

\subsection*{Twin Photon Pair Source Measurements}
For measurements of the on-chip source properties we use a pair of poled waveguides in a domain inversion region with the same parameters as the one in the reconfigurable circuit. This enables direct measurements of source properties without parasitic effects from the subsequent on-chip components. Each source includes an on-chip wavelength division multiplexer (WDM) to separate light around \SI{775}{} (NIR) and \SI{1550}{\nano\meter} (IR). We use cleaved single mode fiber to couple to the sources with monolitically integrated grating couplers designed for the respective wavelength. For the second harmonic measurement in \cref{fig2}a we sweep a tunable CW laser around \SI{1550}{\nano\meter} and synchronously measure at the NIR output port with a power meter. An additional off-chip WDM is used for filtering out any residual pump. For the SPDC measurements we couple a \SI{775}{\nano\meter} CW laser through the NIR port of the on-chip WDM and collect the photon pairs through the IR gratings. We use a fiber based long-pass filter for pump rejection filtering, a 50:50 fiber beamsplitter for probabilistic splitting of the photon pairs, and two superconducting nanowire single photon detectors (SNSPDs) to measure the photon counts. A time tagging unit is used to measure correlations between the signal and idler ports. Additionally, we use manual polarization controllers to control the polarization of the input light as well as the output to optimize the SNSPD detection efficiency. As described in the main text, for some measurements a fiber-based bandpass filter is added before splitting the photons.

The photon spectra in \cref{fig2}c are measured using time-of-flight single photon spectroscopy~\cite{avenhausFiberassistedSinglephotonSpectrograph2009}. A dispersion compensation module with a time dispersion of \SI{0.5}{\nano\second\per\nano\meter} is added in the idler path after the beamsplitter, which introduces a wavelength dependent delay to the idler photon. This stretches the temporal correlation histogram. Using the known bandpass spectrum and dispersion characteristics of the module, the spectral properties of the idler photon is reconstructed.

\subsection*{Bell state generator measurements}
For measurements with the full device, we use a cleaved single mode fiber to couple light at \SI{775}{\nano\meter} through the input grating. The polarization is controlled using an off-chip polarization controller optimized for maximum signal. For collecting the signal we use an 8 channel fiber array, allowing us to couple all four outputs simultaneously. They are subsequently connected to four independent SNSPDs, each with an additional polarization controller to maximize the detection efficiency.  Two bandpass filters are used on the outputs of $Q_A$, which effectively filters $Q_B$ as well since we post-select on measurements with one photon per qubit. These filtering increases the photon coherence length and reduces the impact of small path length mismatches between the two qubit rails. A time tagging unit is used to measure the correlation histogram between all combinations of the output ports. The TO phase shifters are electrically controlled using a programmable multi-channel voltage source. The phase-voltage relationship is modeled as
\begin{equation*}
	\xi(V) = \xi_0 + \frac{\alpha V^2}{1 + \beta V^2},
\end{equation*}
where the random phase offset $\xi_0$ and the parameters $\alpha$ and $\beta$ are obtained through fitting of calibration curves for each individual phase shifter. This relationship is inverted to get the voltage required to apply a target phase. 

\subsection*{Maximum Likelihood Estimation}

To reconstruct the two-qubit quantum states from measured data, we employed maximum likelihood estimation of the density matrix $\rho$~\cite{silverstoneQubitEntanglementRingresonator2015,altepeter4QubitQuantum2004}. The likelihood function $\mathcal{L}(\rho)$ was defined based on the least squares difference between observed coincidence counts and theoretical expectation value of the set of projection measurements. To enforce the physical constraints of $\rho$, we parametrized it as $\rho = T^\dagger T / Tr(T^\dagger T)$, where $T$ is a lower-triangular complex matrix~\cite{altepeter4QubitQuantum2004}. In addition to the $16$ free parameters defining $T$, we also use four normalization paramters capturing the efficiency difference between the SNSPDs used. $\log{\left[ \mathcal{L} ( \rho) \right]}$ was then minimized numerically over the 20 free parameters using a global optimizer. The density matrix $\rho_\text{min}$ minimizing the likelihood function was used to compute relevant figures of merit such as fidelity and concurrence. Statistical uncertainties are estimated through Monte Carlo resampling of the experimentally obtained counts assuming Poissonian distribution. The entire procedure is described in more detail in the supplementary material.

\subsection*{Data availability}
Raw data and evaluation code are available from the authors upon reasonable request.

\subsection*{Competing interests}
The authors declare no competing financial or non-financial interests.

\subsection*{Author contributions}
A.M. and R.J.C. conceived the experiment and designed the photonic circuit. A.S. and J.K. developed and applied the periodic poling of the sample. A.M., G.F., and A.S. fabricated the device including, lithography, etching, metal deposition and wirebonding. A.M. and J.K. performed classical characterization of building blocks. A.M. performed the single photon experiments and with support from R.J.C. performed the data analysis. R.G. supervised the project. A.M. wrote the initial draft of the manuscript. All authors contributed to revising and validating the manuscript content.

\subsection*{Acknowledgements}
We acknowledge support for characterization of our samples from the Scientific Center of Optical and Electron Microscopy ScopeM and from the cleanroom facilities BRNC and FIRST of ETH Zurich and IBM Ruschlikon. 
R.J.C. acknowledges support from the Swiss National Science Foundation under the Ambizione Fellowship Program (Project Number 208707).
R.G. acknowledges support from the European Space Agency (Project Numbers 4000137426 and 4000136423), the Swiss National Science Foundation under the Bridge Program (Project Number 194693) and the Sinergia Program (Project Number CRSII5\_206008).


%apsrev4-2.bst 2015-08-30 from 4.21a (PWD, AO, DPC/HNN) hacked

\clearpage

\renewcommand\thefigure{S\arabic{figure}}  

\setcounter{figure}{0}

\onecolumngrid

\makeatletter
\renewcommand\@biblabel[1]{[S#1]}
\renewcommand\@cite[1]{[S#1]}
\makeatother

\section*{Supplementary Material}
\subsection*{Derivation of On-Chip State Generation}

Here, we discuss the generation and evolution of the state generated by the circuit shown in fig. 1a of the main manuscript.

First, the input power $P_0$ is distributed by the Mach-Zehnder interferometer (MZI) controlled by the phase $\varphi_1$ into two modes $A$ and $B$. The resulting optical power in each mode is
\begin{align*}
	P_A &= P_0 \sin^2{\left( \frac{\varphi_1}{2} \right)} \\
	P_B &= P_0 \cos^2{\left( \frac{\varphi_1}{2} \right)}
\end{align*}

This assumes that the two directional couplers used as beamsplitters are behaving nominally with a reflectance $R = 0.5$. Besides any global phases omitted here, the two pump modes may acquire a relative phase $\theta_1$ due to slight variations in the waveguide dimensions.

The entangled state is generated in the two periodically poled waveguides, which each create a two-mode squeezed vacuum state
\begin{align*}
	\ket{\psi_k} = \sum_{n=0}^\infty \frac{\left[ \tanh{\left(r_k \right)} \right]^n}{\cosh{ \left( r_k \right)}} \ket{n_s n_i}_k,
\end{align*}
where $k = A,B$. The squeezing parameter depends on the respective pump power and the spontaneous parametric down-conversion (SPDC) generation efficiency $\eta_k$, namely $r_k = \eta_k \sqrt{P_k}$ and are therefore implicitly controlled by the phase $\varphi_1$. In a low pump power regime, which we can always achieve by choosing $P_0$ sufficiently small, we can expand the infinite series up to first order and neglect any contributions $\mathcal{O}(r_k^2)$. This allows us to write the state generated by the combined two-mode system as
\begin{equation*}
	\ket{\psi} = \ket{\psi}_A \otimes \ket{\psi_B} = \left( e^{i\theta_1}  \ket{0_s 0_i}_A + r_A e^{i\theta_1} \ket{1_s 1_i}_A \right) \otimes \left(  \ket{0_s 0_i}_B + r_B \ket{1_s 1_i}_B \right).
\end{equation*}
The phase $\theta_1$ is inherited by the SPDC photons since this is a coherent process. By expanding the tensor product we can further rewrite the state to
\begin{equation*}
	\ket{\psi} = e^{i\theta_1} \ket{0_s 0_i}_A \ket{0_s 0_i}_B + r_A  e^{i\theta_1} \ket{1_s 1_i}_A \ket{0_s 0_i}_B + r_B \ket{0_s 0_i}_A \ket{1_s 1_i}_B + r_A r_B  e^{i\theta_1} \ket{1_s 1_i}_A \ket{1_s 1_i}_B.
\end{equation*}
Because of the low pump power regime the circuit operates in, we are safe to assume $r_A \cdot r_B \ll r_A,r_B$ and can therefore neglected the last four photon term. Furthermore, the vacuum state in the first term does not leave experimental signatures in coincidence measurements and can be dropped for the discussion here. This gives us the state generated after the periodically poled waveguides as:
\begin{equation*}
	\ket{\psi} \approx  r_A  e^{i\theta_1} \ket{1_s 1_i}_A \ket{0_s 0_i}_B + r_B \ket{0_s 0_i}_A \ket{1_s 1_i}_B.
\end{equation*}

The splitting of modes $A,B$ into the four modes $a,b,c,d$ and the subsequent crossing lead to the generation of the post-selected state
\begin{align*}
	\ket{\psi} &= r_A  e^{i\theta_1} \ket{1}_a \ket{0}_b \ket{1}_c \ket{0}_d + r_B \ket{0}_a \ket{1}_b \ket{0}_c \ket{1}_d \\ &\hat{=} r_A  e^{i\theta_1} \ket{00} + r_B \ket{11},
\end{align*}
where in the final step we used standard two-qubit state notations assuming that modes $a,b$ and $c,d$ each form a dual-rail encoded qubit. This state, up to the phase factor, form an entangled Bell state.

The phase $\theta_2$ introduced by the thermo-optic (TO) phase shifter after the first splitting controls this phase factor and the MZI controlled by $\theta_2$ adjust the state of the second qubit. We can include these phases in our final state:
\begin{align*}
	\ket{\psi} = r_A  e^{i\left(\theta_1 + 2 \theta_2 \right)} \left[ \sin{\left(\frac{\varphi_2}{2} \right)} \ket{00} + \cos{\left(\frac{\varphi_2}{2} \right)} \ket{01} \right] + r_B \left[ \cos{\left(\frac{\varphi_2}{2} \right)} \ket{10} - \sin{\left(\frac{\varphi_2}{2} \right)} \ket{11} \right].
\end{align*}

Finally, to write the full dependence on TO controlled phases, we can include the implicit dependence on $\varphi_1$ through the squeezing parameters:
\begin{align*}
	\ket{\psi} = \qquad \eta_a e^{i \left( \theta_1 + 2\theta_2 \right)} \sin{\left( \frac{\varphi_1}{2} \right)} \sin{\left( \frac{\varphi_2}{2} \right)} &\ket{00} \\ +~  \eta_a e^{i \left( \theta_1 + 2\theta_2 \right)} \sin{\left( \frac{\varphi_1}{2} \right)} \cos{\left( \frac{\varphi_2}{2} \right)}  &\ket{01} \\ + \eta_b \cos{\left( \frac{\varphi_1}{2} \right)} \cos{\left( \frac{\varphi_2}{2} \right)} &\ket{10} \\ - \eta_b  \cos{\left( \frac{\varphi_1}{2} \right)} \sin{\left( \frac{\varphi_2}{2} \right)} &\ket{11}.
\end{align*}
Note that we set $P_0 = 1$ since it is just a normalization constant. Also, we assumed perfect 50:50 beamsplitters throughout this derivation. In the case of pumping with equal strengths, that is $\varphi_1 = \pi/2$, and assuming $\eta_a = \eta_b \equiv \eta$ we can rewrite the computational basis states as linear combinations of the Bell states, for example $\ket{00} = \left( \ket{\Phi^+} + \ket{\Phi^-}\right / \sqrt{2}$, to arrive at an expression of the state in Bell basis:
\begin{align*}
	\ket{\psi} = \qquad \frac{\eta}{\sqrt{2}} \left(  e^{i \left( \theta_1 + 2\theta_2 \right)} - 1 \right) \sin{\left( \frac{\varphi_2}{2} \right)} &\ket{\Phi^+} \\ + \frac{\eta}{\sqrt{2}} \left(  e^{i \left( \theta_1 + 2\theta_2 \right)} + 1 \right) \sin{\left( \frac{\varphi_2}{2} \right)} &\ket{\Phi^-} \\
	+ \frac{\eta}{\sqrt{2}} \left(  e^{i \left( \theta_1 + 2\theta_2 \right)} + 1 \right) \cos{\left( \frac{\varphi_2}{2} \right)} &\ket{\Psi^+} \\
	+ \frac{\eta}{\sqrt{2}} \left(  e^{i \left( \theta_1 + 2\theta_2 \right)} - 1 \right) \cos{\left( \frac{\varphi_2}{2} \right)} &\ket{\Psi^-}
\end{align*}

By further simplifying by setting $\theta_1 = 0$ (which can always be achieved by shifting the phase setting of $\theta_2$ by $\theta_1/2$) and subsequently rewriting the exponentials as trigonometric functions, we arrive at \cref{eq:BellBasis} in the main manuscript for the state.

\subsection*{Maximum Likelihood Estimation}

Reconstructing the full density matrix of a quantum state theoretically requires 16 measurements projecting the state onto different basis~\cite{altepeter4QubitQuantum2004}. Enabled by simultaneous coupling and measurement of both qubit rails with four independent single photon detectors, projections along the positive and negative direction of an axis is implemented without additional effort. Furthermore, the high pair generation rates of our on-chip sources lead to sufficiently short integration times, such that projecting onto all combinations of $X$, $Y$ and $Z$ axes of the Bloch sphere for each qubit does not significantly increase the measurement duration. Therefore, we use the full set of 36 projection measurements to perform the density matrix reconstruction. This has been shown to improve performance when using maximum likelihood estimation~\cite{deBurgh2008}.

Experimentally, we perform full two-qubit tomography by implementing all pairs of local Pauli operators $\Pi_A, \Pi_B \in \left\{ \sigma_x, \sigma_y, \sigma_z \right\}$. Each of the 9 combinations $(\Pi_A, \Pi_B)$ is implementing using a specific phase setting $\Xi_\ell = (\varphi_{3,\ell}, \theta_{3,\ell}, \varphi_{4,\ell}, \theta_{4,\ell})$. For a given $\Xi_\ell$ we set the projection phases and record the coincidence counts $C_{jk,\ell}$ between output rails $j \in \left\{ a,b\right\}$ of $Q_A$ and $k \in \left\{ c,d\right\}$ of $Q_B$. The values for $C_{jk,\ell}$ are obtained by summing the correlation histogram and subtracting accidental counts. Depending on which the specific combination of qubit modes, a different combination of projections onto $(P_A, P_B)$ is measured (see Fig. 3c of the main manuscript). 

Since the measurements use four independent detectors, care has to be taken to consider varying detection efficiencies. Lumped into detection efficiency are also variations in coupling efficiencies or transmission losses in fiber filters. Therefore, we use a slightly adjusted version of the standard MLE procedure detailed here.

The likelihood function used for MLE is
\begin{equation}
	\mathcal{L}(\hat{\rho}) = \sum_{j = 1}^{9}  \sum_{\substack{j \in \left\{ a,b\right\} \\ k \in \left\{ c,d\right\}}} \left\vert C_{jk,\ell} - \tilde{C}(\hat{\rho}, \Pi_{jk,\ell})\right\vert^2
\end{equation}
where $C_{jk,\ell}$ are the measured coincidences and $\tilde{C}(\hat{\rho}, \Pi_{jk,\ell})$ is the theoretically expected number of coincidences for the state $\hat{\rho}$ under the two-qubit projector $\Pi_{jk,\ell}$. The expected coincidence counts are derived from the theoretical probability $p(\hat{\rho}, \Pi) = \mathrm{Tr}{\left( \hat{\rho} \cdot \Pi \right)}$ by normalizing them via
\begin{equation}
	\tilde{C}(\hat{\rho}, \Pi_{jk,\ell}) = p(\hat{\rho}, \Pi_{jk,\ell})\frac{\sum_{j = 1}^{9}  \sum_{\substack{j \in \left\{ a,b\right\} \\ k \in \left\{ c,d\right\}}} \mathcal{N}_{jk} C_{jk,\ell}}{\sum_{j = 1}^{9}  \sum_{\substack{j \in \left\{ a,b\right\} \\ k \in \left\{ c,d\right\}}} p(\hat{\rho}, \Pi_{jk,\ell})}.
\end{equation}
Here, we introduced additional normalization constants $\mathcal{N}_{jk}$ which take varying detection efficiencies for the different combinations of qubit modes into account. The density matrix $\hat{\rho}$ is parametrized as $\rho = T^\dagger T / Tr(T^\dagger T)$, where $T$ is a lower-triangular matrix parametrized by $16$ real parameters and we use the four normalization constant as additional free parameters. We numerically minimize $\log{\left[ \mathcal{L}(\hat{\rho})\right] }$ using the L-BFGS-B algorithm implemented in the SciPy package. The result of this minimization is the density matrix $\hat{\rho}_0$ which best describes the measurement results.

\subsection*{Full Experimental Results for State Preparation}

\begin{figure*}[t!]
	\includegraphics[width=\textwidth]{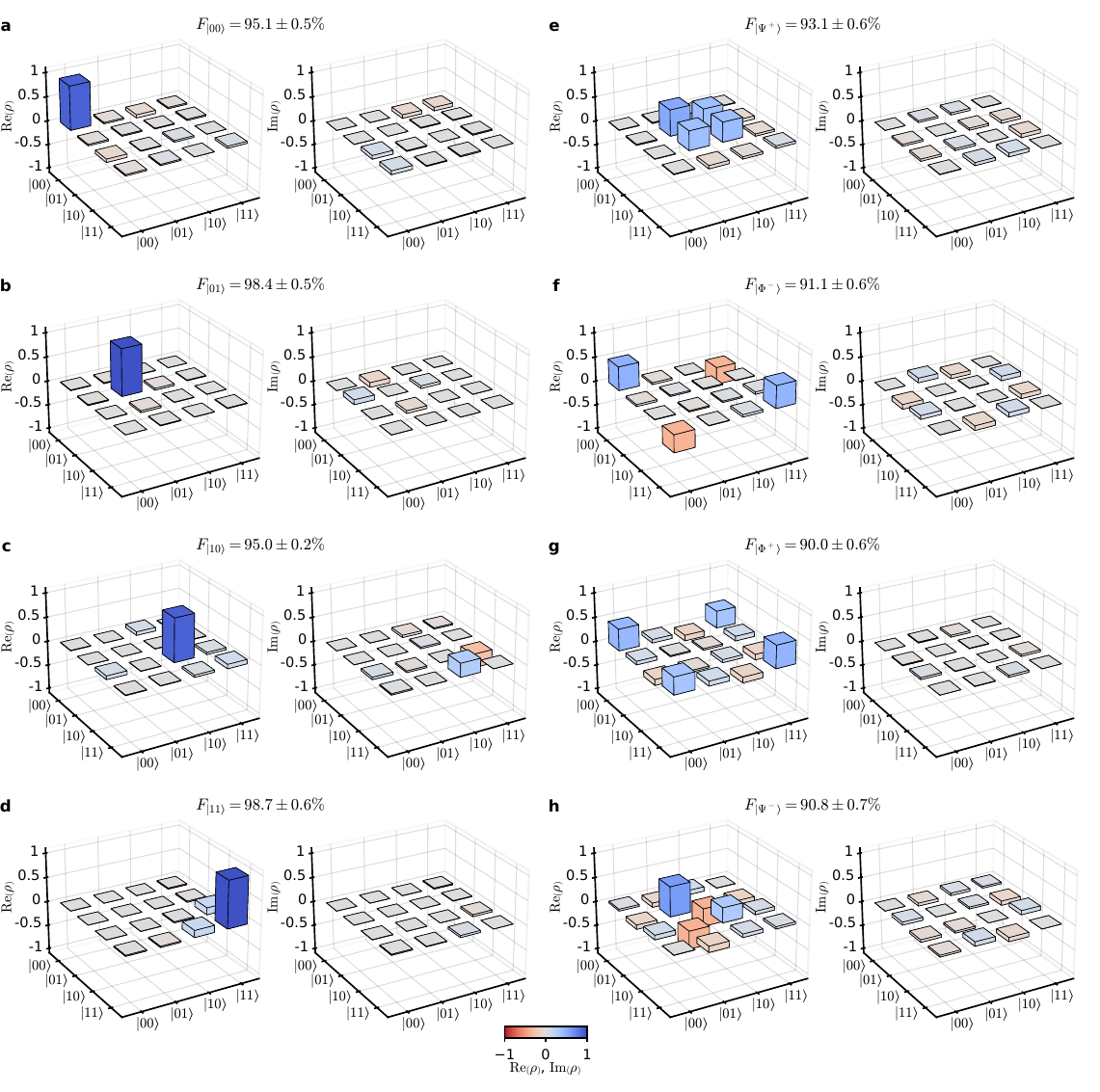}
	\caption{\textbf{Full Reconstructed Density Matrices.} Real and imaginary parts of the density matrices obtained via maximum likelihood estimation (MLE) for the states (\textbf{a}) $\ket{00}$, (\textbf{b}) $\ket{01}$, (\textbf{c}) $\ket{10}$, (\textbf{d}) $\ket{11}$, (\textbf{e}) $\ket{\Psi^+}$, (\textbf{f}) $\ket{\Psi^-}$, (\textbf{g}) $\ket{\Phi^+}$, and (\textbf{h}) $\ket{\Phi^-}$. These plots provide a complete representation of the reconstructed two-qubit states, complementing the real-part-only visualizations shown in the main text. Fidelities $F_{\ket{\psi}}$ refer to the overlap between the reconstructed density matrix and the respective target state $\ket{\psi}$.} \label{figS1}
\end{figure*}

The imaginary components of the reconstructed density matrices for the states presented in the main manuscript are shown in \cref{figS1} for completeness. For each state, we also provide the fidelity $F$, concurrence $C$, and the von Neumann entropies $S_A$ and $S_B$ of the reduced density matrices obtained by tracing out one qubit, as summarized in \cref{tab1}. All metrics are computed using the QuTiP library in Python. Uncertainties are estimated as the standard deviation from Monte Carlo resampling of the experimentally measured counts, assuming Poissonian statistics, followed by maximum likelihood estimation for each sample.

\begin{table*}[tbp!]
	\caption{\label{tab:ex} Summary of the experimental results for the generation of target quantum states. The table lists the input phase settings $\varphi_1$, $\varphi_2$ and $\theta_2$ used in the state preparation, along with the fidelity $F$, concurrence $C$, and von Neumann entropies $S_A$ and $S_B$ of the reduced density matrices obtained by tracing out one qubit. All values include one standard deviation uncertainty, estimated via Monte Carlo resampling.} \label{tab1}
	\begin{ruledtabular}
		\begin{tabular}{lccccccc}
			Target State &  $\varphi_1$ & $\varphi_2$ & $\theta_2$ & $F$ [\SI{}{\percent}] & $C$ & $S_A$ & $S_B$ \\ \hline
			$\ket{00}$ & $\pi$ & $0$ & $\pi$ & $95.1 \pm 0.5$ & $0.13 \pm 0.03$ & $0.25 \pm 0.02$ & $0.23 \pm 0.2$ \\
			$\ket{01}$ & $\pi$ & $0$ & $0$ & $98.4 \pm 0.5$ & $0.13 \pm 0.02$ & $0.09 \pm 0.04$ & $0.05 \pm 0.02$ \\
			$\ket{10}$ & $0$ & $0$ & $\pi$ & $95.0 \pm 0.2$ & $0.05 \pm 0.01$ & $0.006 \pm 0.004$ & $0.006 \pm 0.003$ \\
			$\ket{11}$ & $0$ & $0$ & $0$ & $98.7 \pm 0.6$ & $0.06 \pm 0.02$ & $0.017 \pm 0.005$ & $0.014 \pm 0.004$ \\
			$\ket{\Psi^+}$ & $\pi/3$ & $0$ & $0$ & $93.1 \pm 0.6$ & $0.8 \pm 0.03$ & $0.682 \pm 0.005$ & $0.67 \pm 0.01$ \\
			$\ket{\Psi^-}$ & $\pi/3$ & $0$ & $\pi/2$ & $90.8 \pm 0.7$ & $0.8 \pm 0.02$ & $0.64 \pm 0.01$ & $0.64 \pm 0.01$ \\
			$\ket{\Phi^+}$ & $\pi/3$ & $\pi$ & $\pi/2$ & $90.0 \pm 0.6$ & $0.73 \pm 0.02$ & $0.688 \pm 0.004$ & $0.689 \pm 0.003$ \\
			$\ket{\Phi^-}$ & $\pi/3$ & $\pi$ & $0$ & $91.1 \pm 0.6$ & $0.74 \pm 0.02$ & $0.692 \pm 0.001$ & $0.693 \pm 0.001$ \\
		\end{tabular}
	\end{ruledtabular}
\end{table*}

\FloatBarrier

\end{document}